\documentclass[11pt,a4paper,oneside]{article}
\usepackage[top=3cm, bottom=3cm, left=2cm, right=2cm]{geometry}
\usepackage[english]{babel}
\usepackage[utf8x]{inputenc}
\usepackage[T1]{fontenc}

\usepackage{amsthm,amsmath,amssymb,mathrsfs,dsfont,mathtools}

\usepackage{bm}
\usepackage{graphicx}
\usepackage[colorinlistoftodos]{todonotes}
\usepackage[colorlinks=true, allcolors=blue]{hyperref}
\usepackage{booktabs,subcaption}
\usepackage{hyperref}
\usepackage[all]{nowidow}
\usepackage{setspace}
\usepackage{algpseudocode}
\usepackage{algorithm}
\usepackage{tikz} 
\usepackage[sf]{titlesec}
\usepackage{sectsty}
\allsectionsfont{\centering \normalfont\scshape} 
\usepackage{lipsum}
\usepackage{adjustbox}
\usepackage{lineno} 
\usepackage{natbib}

\usepackage{authblk}
\title{Inferring taxonomic placement from DNA barcoding allowing discovery of new taxa}
\date{}
\author[1]{Alessandro Zito}
\author[2]{Tommaso Rigon} 
\author[1]{David B. Dunson}

\affil[1]{Department of Statistical Science, Duke University, Durham, NC, 27708, U.S.A.}
\affil[2]{Department of Economics, Management and Statistics, University of Milano-Bicocca, Milan, 20126, Italy}

\theoremstyle{definition}

\newcommand{\bayesant}{BayesANT }
\newcommand{\bV}{\mathbf{V}}
\newcommand{\bX}{\mathbf{X}}
\newcommand{\pa}{\text{pa}}
\newcommand{\btheta}{\boldsymbol{\theta}}
\newcommand{\K}{\mathcal{K}}
\newcommand{\kmer}{$\kappa$-mer~}
\newcommand{\kmers}{$\kappa$-mers~}

\newcommand{\bxi}{\boldsymbol{\xi}}

\begin{document}

\maketitle

\begin{abstract}
In ecology it has become common to apply DNA barcoding to biological samples leading to datasets containing a large number of nucleotide sequences.  The focus is then on inferring the taxonomic placement of each of these sequences by leveraging on existing databases containing reference sequences having known taxa.  This is highly challenging because i) sequencing is typically only available for a relatively small region of the genome due to cost considerations; ii) many of the sequences are from organisms that are either unknown to science or for which there are no reference sequences available.  These issues can lead to substantial classification uncertainty, particularly in inferring new taxa.  To address these challenges, we propose a new class of Bayesian nonparametric taxonomic classifiers, BayesANT, which use species sampling model priors to allow new taxa to be discovered at each taxonomic rank.  Using a simple product multinomial likelihood with conjugate Dirichlet priors at the lowest rank, a highly efficient algorithm is developed to provide a probabilistic prediction of the taxa placement of each sequence at each rank. BayesANT is shown to have excellent performance in real data, including when many sequences in the test set belong to taxa unobserved in training. 
\end{abstract}

\section{Introduction}

DNA barcoding refers to the practice of identifying the taxonomic affiliation of unknown specimens through short fragments of their DNA molecular sequence called barcoding genes. Typically, this assessment is performed by comparing the DNA obtained from the high-throughput sequencing of a bulk sample to a  library of genes whose Linnean taxonomy is well-established. Examples of these collections are numerous, with the Barcode of Life project \citep{Sarkar_Trizna_2011} being a particularly notable case.  In order for the identification to be reliable, reference DNA sequences should be characterized by limited intra-species and high inter-species gene variation, and should be sufficiently simple to align and compare \citep{Hebert_2003}. In the animal kingdom and in insects especially, these characteristics have been found in a region of approximately 650-base-pairs near the 5th end of the mitocondrial cytochrome c oxidase sub-unit I, or COI, gene \citep{Jazen_2005}.  This region has become routinely used in animal species identification.
In particular, libraries are formed by clustering similar COI sequences under a common Barcode Index Number, or BIN, which identifies a given species \citep{Ratnasingham_2013}.

The impact of DNA barcoding in biodiversity assessment 
has been dramatic. It took more than 200 years to describe approximately 1 million species of insects through morphological inspection, whereas nearly 400,000 BINs have been categorized within the span of just 10 years \citep{Wilson_2017}. DNA barcoding offers a way to categorize the large quantities of specimens collected by modern automatic sampling methods. For example, flying insects are routinely captured with Malaise traps \citep{Malaise_1937},
which collect the sampled insects together in a `soup' within a storage cylinder.
While this method often causes deterioration of the captured animals, making them morphologically unrecognizable, the biologic material in the soup can be processed relatively cheaply \citep{Shokralla_2014} and later used for identification. The practice of DNA sequencing from a bulk sample is known as DNA metabarcoding \citep{Yu_2012}. The output of a metabarcoding procedure is the grouping of similar sequences detected in the samples into \textit{operational taxonomic units}, or OTUs. These OTUs provide initial hypothesized species labels for the animals in the sample, and assessing their taxonomic placement is the final key stage of a bioinformatics pipeline.

Despite the advantages described above, taxonomic assessment of OTUs presents its own challenges, especially at lower level ranks. While it is relatively easy to accurately place a DNA sequence to a \textit{Phylum}, a \textit{Class} or an \textit{Order} \citep{Yu_2012}, the information obtainable via high-throughput methods is limited by the short length of the sequences extracted. This makes the identification at the \textit{Family}, at the \textit{Genus} and at the \textit{Species} level subject to higher uncertainty \citep{Pentinsaari_2020}. Moreover, DNA metabarcoding may be prone to sequencing and clustering errors. As a consequence, it can either split biologic material from the same species into two different clusters, or merge different species into a single cluster \citep{Somuervuo_2017}. 
Finally, reference sequence libraries can be subject to mislabelling errors \citep{Somervuo_2016} and are necessarily incomplete. 
This leads to the necessity of developing classification methods that provide a reliable characterization of uncertainty in assigning the taxa of the collected OTUs, 
accounting for the possibility that they might identify a new species or one for which information is not available. Such methodologies 
allow one to quantify the biodiversity of a given sampling region, 
which is an urgent problem considering the recent evidence of insect biomass decline \citep{Seibold_2019}.

Various software for taxonomic recognition have been developed, relying on different prediction methods. One approach labels a query DNA with the taxon of the reference sequence having the highest similarity \citep{Huson_2007, Nguyen_2014}. This requires applying local or global alignment procedures to the sequences in the library, such as the BLAST - Basic Local Alignment Search Tool - similarity score \citep{Altschul1990}. When alignment is undesirable due to computational costs, fast algorithms that exploit a ${\kappa}$-mer representation of the sequences can be adopted. Widely used examples are the Na\"{i}ve Bayes RDP classifier \citep{Wang_2007} and its non-Bayesian heuristic alternatives  \citep[e.g SINTAX,][]{Edgar_2013}. More recent methods use modern Machine Learning and Deep Learning techniques, such as Convolutional Neural Networks \citep{Vu_2020}.

While these approaches can provide good classification results when the training data are sufficiently informative of the biodiversity of the environmental sample \citep{Bazinet_2012}, they can lead to unreliable matches when the reference sequence set is incomplete.
Indeed, algorithms often do not account for unobserved taxa in a coherent way \citep{Somuervuo_2017}. One solution is to select a confidence probability cutoff, typically around 0.8, and regard the classification as unreliable if the predicted taxon has a probability below that threshold \citep{Wang_2007, Lan_2012}. Another alternative is to explicitly allow the algorithm to predict ``new'' taxa, as is done by PROTAX - PRObabilistic TAXonomic placement \citep{Somervuo_2016}. PROTAX classifies DNA sequences by training a multinomial regression model on a subsample of the reference library reflecting prior knowledge on the existing taxonomy. The algorithm can lead to over- or under-detection of new species if the training dataset is not representative.


In this paper, we develop an off-the-shelf Bayesian nonparametric  model for DNA barcoding that accounts for undetected nodes at every taxonomic rank. As our application mostly focuses on insects, we name our method BayesANT, short for BAYESiAn Nonparametric Taxonomic classifier. BayesANT computes taxon probabilities by combining a prior distribution for the taxonomic tree with a kernel-based approach to modelling the distribution of the nucleotide sequences linked to that taxon at the lowest rank.
To obtain the prior, a Pitman--Yor process \citep{Perman_1992} is used to induce a species sampling model urn scheme \citep{Blackwell1973, Pitman1996}, which automatically specifies probabilities for the appearance of undiscovered species   
\citep{Lijoi_mena_pruenster_2007, Favaro_lijoi_mena_pruenster_2009, Zito_2020} in a coherent way.
For aligned sequences, we use a Dirichlet-multinomial product kernel over nucleotides, while, for unaligned sequences, we use a multinomial kernel over $\kappa$-mer counts.
The resulting model facilitates fast computation of a probabilistic classifier, which provides careful uncertainty assessments in taxonomic annotations. We test BayesANT on a library of arthropod DNA sequences collected in Finland \citep{Roslin_2021}. Technical and algorithmic details for BayesANT are reported in the Materials and Methods section, with further information in the Supplementary material. 

\section{Preliminaries: the Pitman--Yor process} 

\begin{figure}[tb]
    \centering
    \includegraphics[width=0.7\textwidth]{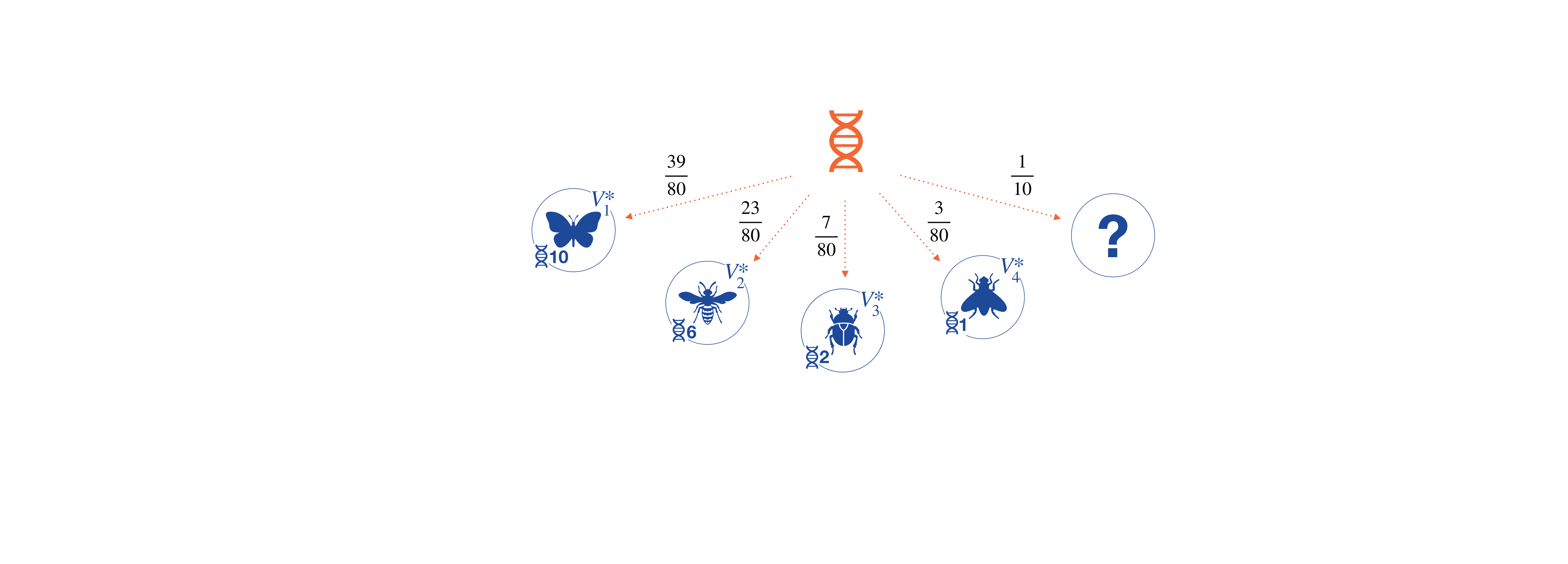}
    \caption{\small{Example of a Pitman--Yor process with $n= 19$, $\alpha = 1$, $\sigma = 0.25$ and $K_n=4$. Taxa names are reported on top of the circles, and frequencies of appearance are written on the right to the blue DNA sequences, respectively. Fractions in black denote the taxon probabilities for the orange DNA sequence. For example, the probability of observing the butterfly-shaped taxon $V_1^*$ is $(n_1 - \sigma)/(\alpha+n) = (10 - 0.25)/(19 + 1) = 39/80$. The probability for the unknown question mark taxon is $(\alpha + \sigma k)/(\alpha+n) = (1 + 4\times0.25)/(19 + 1) = 1/10$.}}
    \label{fig:Pitman-Yor}
\end{figure}

The Pitman--Yor \citep{Perman_1992} is a sequential process for label assignment whose allocation probabilities depend on two parameters, called  $\alpha$ and $\sigma$, and on the size of the clusters previously detected. The allocation rule works as follows. Suppose that $V_1, \dots, V_n$ are the taxon assignments for the DNA sequences in our library of barcodes at a given rank (such as \textit{Phylum}  or \textit{Class}). Specifically, these sequences identify $K_n=k$ distinct taxa, named $V_1^*, \ldots, V_k^*$, with frequencies $n_1, \ldots, n_k$ and $\sum_{j=1}^k n_j = n$. Then, the probability that the $(n+1)$st sequence belongs to the $j$th of the known taxa is 
\begin{equation}\label{eq:PY_probOld}
    \text{pr}(V_{n+1} = V_j^* \mid V_1, \ldots, V_n) = \frac{n_j - \sigma}{\alpha + n},
\end{equation}
for $j = 1, \ldots, k$, while the probability of observing a new taxon is 
\begin{equation}\label{eq:PY_probNew}
    \text{pr}(V_{n+1} = \text{``new''} \mid V_1, \ldots, V_n) = \frac{\alpha + \sigma k}{\alpha + n},
\end{equation}
where $\alpha > - \sigma$ and $\sigma \in [0,1)$. Figure~\ref{fig:Pitman-Yor} sketches the mechanism when $n=19$ sequences and $k=4$ different groups are observed.
High values of $\alpha$ or values of $\sigma$ close to 1 lead to a high probability of discovering a new taxon.  The probability that a sequence is assigned to taxon label $V_j^*$ increases with its abundance $n_j$.  
This process allows   
 OTUs to be clustered together, through being assigned to the same existing or newly detected taxa.  For further details, see \citep{Pitman1996, DeBlasi2015}. 


\section{Case study: the Finnish Barcode of Life library}



\subsection{The FinBOL library}

The Finland Barcode of Life initiative\footnote{ \hyperlink{https://en.finbol.org/}{https://en.finbol.org/}} (FinBOL) is a DNA barcoding library that contains reference sequences with highly reliable taxonomic annotations for the arthropod species of Finland. The data have been constructed placing substantial effort on barcode quality thanks to the collective effort of about 150 taxonomists. For a thorough description of how the library was assembled and later tested refer to \citet{Roslin_2021}.

The version of the data we consider contain a total of $34,624$ DNA sequences annotated across 7 taxonomic levels, namely \textit{Class}, \textit{Order}, \textit{Family}, \textit{Subfamily}, \textit{Tribe}, \textit{Genus} and \textit{Species}. Reference annotations are based off of the national checklist of Finnish species \citep{FinBIF_2020} with the inclusion of dummy taxa whenever \textit{Subfamily} and \textit{Tribe} were missing. The library has been globally aligned via Hidden Markov Models using the HMMER software \citep{Eddy_1995}. As a result, each sequence has a length of $658$ base pairs, consisting of nucleotides ``A'', ``C'', ``G'' and ``T'' and alignment gaps ``-''. Other special characters are ignored and substituted with a gap for simplicity. Taxonomic labels in the data comprise 3 \textit{Classes}:  \textit{Arachnida},  \textit{Insecta} and  \textit{Malacostraca}, appearing 1,842 and 32,781  and 1 times, respectively. The sequences are further divided into 21 \textit{Orders}, 476 \textit{Families}, 896 \textit{Subfamilies}, 1,355 \textit{Tribes}, 3,855 \textit{Genera} and 10,985 \textit{Species}, 3,025 of which have a single reference sequence associated to them.

\begin{figure}[tb]
    \centering
    \includegraphics[width=0.5\textwidth]{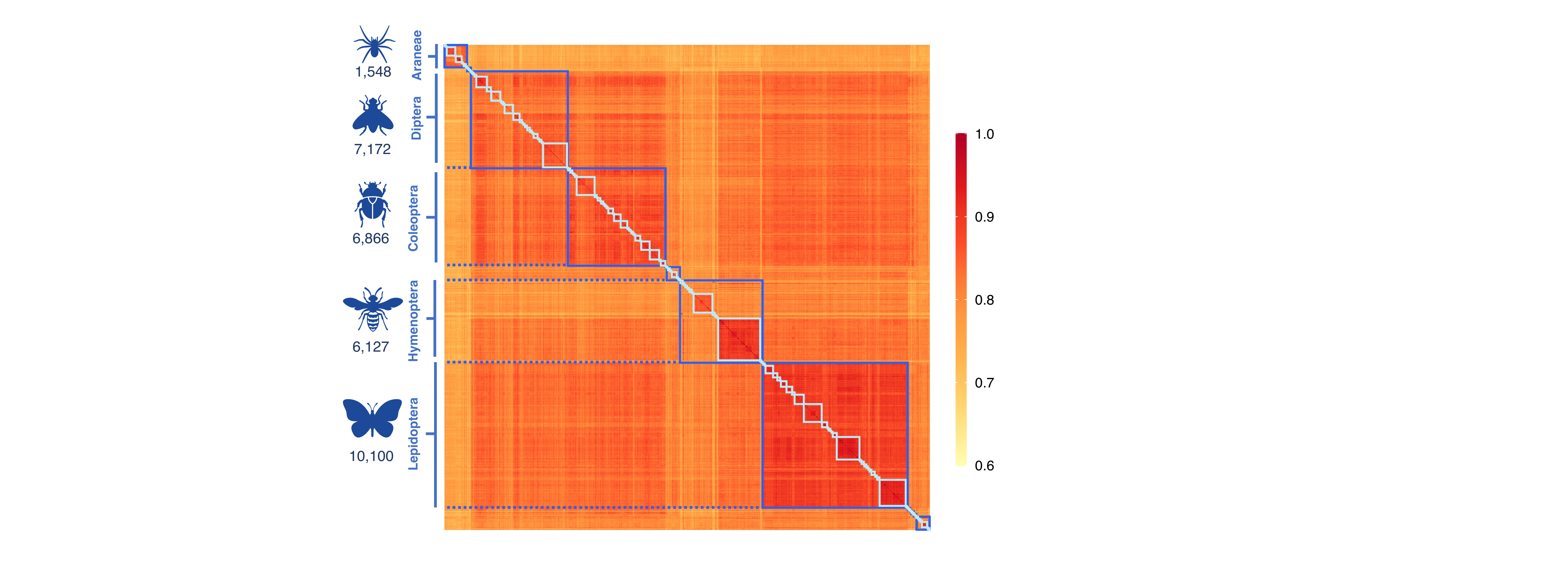}
    \caption{\small{Raw DNA similarities between 3,000 randomly sampled sequences from the FinBOL library. The blue and light blue boxes along the main diagonal identify the \textit{Orders} and the \textit{Families}, respectively. Numbers on the left side represent the frequencies for the 5 largest \textit{Orders} in the data.}}
    \label{fig:FINBOL_similarities}
\end{figure}

Figure~\ref{fig:FINBOL_similarities} depicts the pairwise raw DNA similarities, calculated as the fraction of locations with equal nucleotides, between 3,000 randomly sampled sequences from the library. Each row/column represents the  DNA similarity between one sequence and all the other sampled ones, with darker tones indicating higher similarities. Sequences are sorted according to the alphabetical order of their annotation to ensure a cluster separation. In particular, boxes in dark blue along the main diagonal highlight the cross similarities within the \textit{Orders}, while boxes in light blue refer to the \textit{Families}. On the left side of the Figure we report the name and the sizes of the 5 most frequent orders, namely \textit{Araneae}, \textit{Diptera}, \textit{Coleoptera}, \textit{Hymenotptera} and \textit{Lepidoptera}. In an ideal setting, the within-taxon similarities along the main diagonal should be higher than the cross-taxa ones. However, this is only true for \textit{Lepidoptera} and for the two largest families - \textit{Ichneumonidae} and \textit{Tenthredinidae} - in \textit{Hymenoptera}. Indeed, \textit{Diptera} and \textit{Coleoptera} are virtually indistinguishable, as they show a similar within- and between-order similarity. Moreover, these two taxa show a high cross-similarity with \textit{Lepidoptera}, as indicated by the off-diagonal orange rectangles.  Overall, the average DNA similarity in the library is around 0.81, indicating that the sequences are highly homogeneous. 

\subsection{Testing scenarios}
\begin{figure}[tb]
    \centering
    \includegraphics[width=0.6\textwidth]{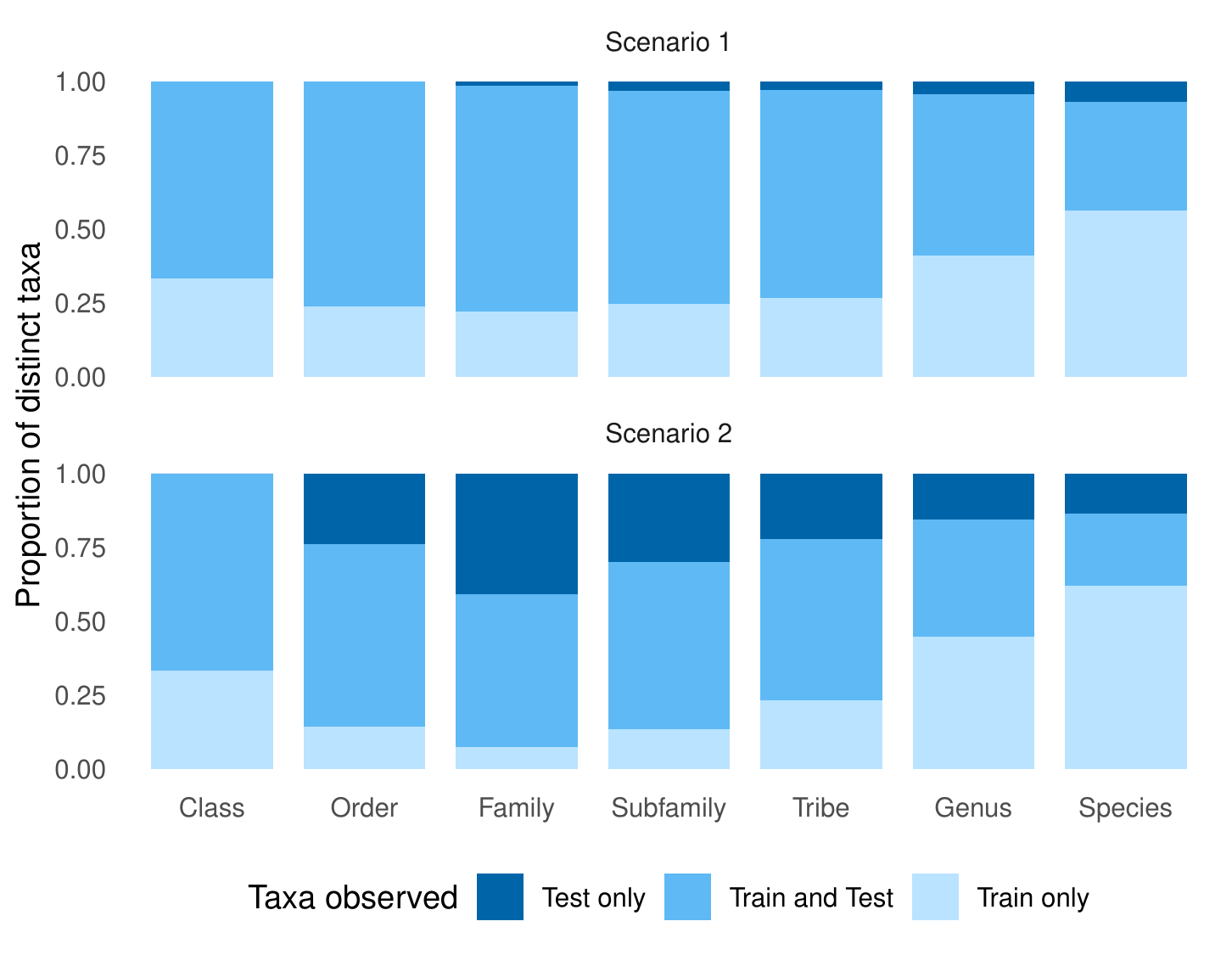}
    \caption{\small{Taxonomic composition of the training and the test libraries in the two splitting scenarios.}}
    \label{fig:FINBOL_splits}
\end{figure}

Our aim is to evaluate the performance of the predictive taxa classification probabilities produced by BayesANT. If a sequence is from an existing taxon at a certain rank, that taxon should be assigned relatively high probability.  If the sequence is instead from a new taxa, the predictive probabilities should reflect this reality.  We train the classifiers on a random subset of 80\% of the FinBOL data and predict the taxonomic affiliation for the remaining 20\% of the sequences.  Some taxa in the test set will not appear in the training data; most existing methods will automatically misclassify these sequences as belonging to one of the taxa in the training library.

We consider two testing scenarios summarized in Figure~\ref{fig:FINBOL_splits}. In the first, each sequence in the library has equal probability of being allocated to the test set. This makes the taxonomic composition of the training and test set similar. As a result, only a relatively small fraction of the taxa will be unobserved in training, as is evident from the top panel of Figure~\ref{fig:FINBOL_splits}. In the second scenario, we create the test set by stratified sampling: for each test observation, we first sample the \textit{Family}, and then draw one sequence within that \textit{Family}. This assigns each \textit{Family} an equal probability of being selected, irrespective of its frequency of appearance in the data. Such a procedure yields a different composition between training and test, resulting in many more test taxa that are unobserved in training. See Figure~\ref{fig:FINBOL_splits}, bottom panel.

\subsection{Results} 


BayesANT computes the probability of every node in the taxonomic tree, including potential novel ones, for every test DNA sequence. The predicted  annotation is the taxonomic branch with the highest probability at every rank. These probabilities express the uncertainty of the classification, and need to be well calibrated to be reliable: for instance, if 90\% of the sequences are correctly classified, then the average probability with which they are classified should be around 0.9. Ideally, we would like to limit cases in which the algorithm is too confident when wrong and too conservative when right; see the Materials and Methods section. Moreover, evaluating the performance of BayesANT requires a clear definition of correctness of the classification under novel taxa. If the true annotation of a test sequence shows a taxon which is unobserved in training, the prediction outcome may be the correct novel taxonomic leaf, or a new taxon but in an incorrect branch, or a taxon observed in training. We consider the classification to be correct in the first case and wrong otherwise. For example, if the true annotation of test sequence is 
\begin{small}
\begin{verbatim}
    Insecta -> Diptera-> Tephritidae -> Trypetinae -> Trypetini -> Acidia -> Acidia cognata
\end{verbatim}
\end{small}
but \texttt{Acidia} is a \textit{Genus} not observed in the training set, then the correct classification up to the \textit{Species} rank is
\begin{small}
\begin{verbatim}
    Insecta -> Diptera -> Tephritidae -> Trypetinae -> Trypetini -> New Genus in Trypetini 
            -> New Species in New Genus in Trypetini
\end{verbatim}
\end{small}
since the novelty  produces a new \textit{Genus} and automatically a new \textit{Species} linked to it. As \texttt{Acidia} is not observed, necessarily also the \textit{Species} \texttt{Acidia cognata} is unseen and the classification at the  \textit{Species} level is correct only if BayesANT recognizes the novel \textit{Genus}. An outcome such as
\begin{small}
\begin{verbatim}
    Insecta -> Diptera -> Tephritidae -> Trypetinae -> Trypetini -> Trypeta 
            -> New Species in Trypeta
\end{verbatim}
\end{small}
is wrong but recognizes a novel leaf, while 
\begin{small}
\begin{verbatim}
    Insecta -> Diptera -> Tephritidae -> Trypetinae -> Trypetini -> Trypeta -> Trypeta zoe  
\end{verbatim}
\end{small}
is wrong since it predicts an observed \textit{Species}.

\begin{figure}[tb]
    \centering
    \includegraphics[width=0.7\textwidth]{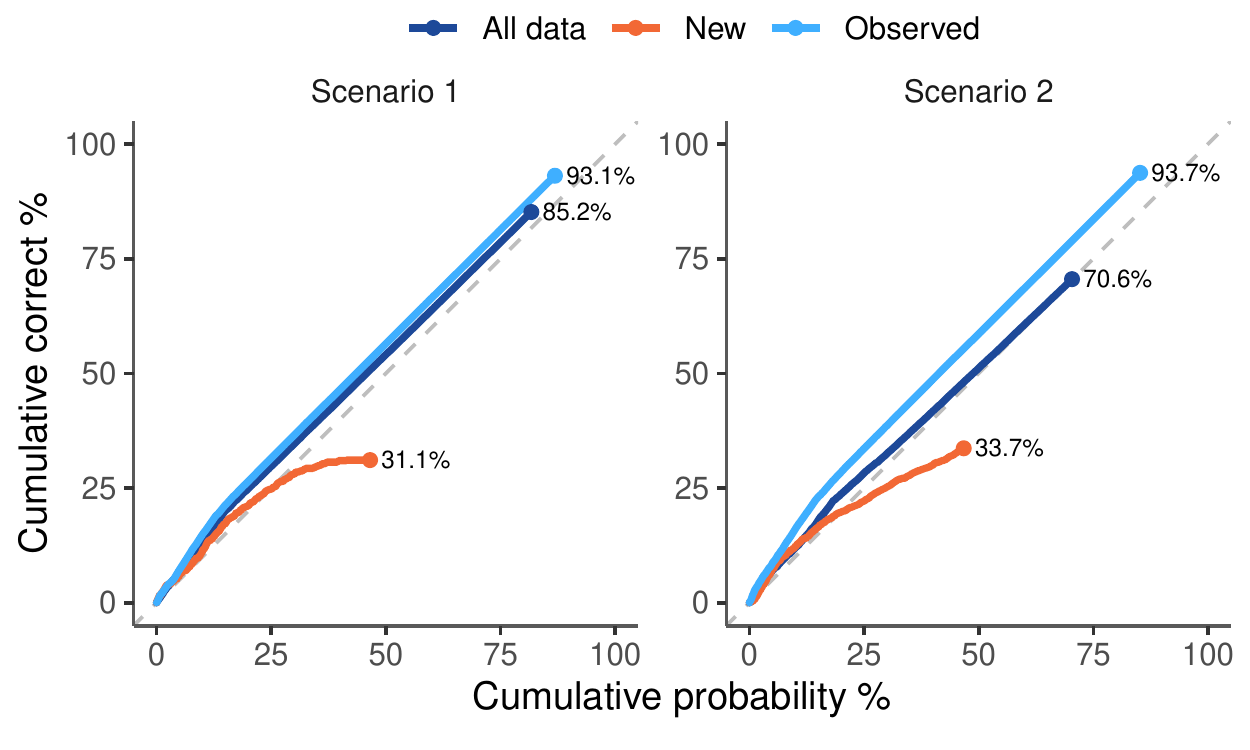}
    \caption{\small{Calibration plot for the prediction of BayesANT at the \textit{Species} level under both scenarios. The dashed diagonal line indicates perfect calibration, while the percentages next to the points are the \textit{Species} accuracies in the test sets. Notice that ``All data'' includes all the query sequences in the test set, ``New''  refers to those whose true taxon is not observed in training at some rank, while ``Observed'' restricts to the cases where the true taxonomy is fully observed. 
    }}
    \label{fig:FINBOL_performances}
\end{figure}

Figure~\ref{fig:FINBOL_performances} displays the prediction probabilities of BayesANT in both FinBOL scenarios. As the library is globally aligned, we adopt a simple product-multinomial kernel in which the probabilities of nucleotides ``A'', ``C'', ``G'' and ``T'' vary by loci and species. We treat the alignment gap ``-'' as a missing value and ignore the likelihood contribution of the locations where it appears. The rank-specific parameters $\alpha_\ell$ and $\sigma_\ell$ are estimated from the data as we detail in the Supplementary material.  Figure~\ref{fig:FINBOL_performances} depicts the relationship between the \% cumulative probability and the \% cumulative accuracy at the \textit{Species} level. 
The dashed diagonal indicates a perfectly calibrated output, while trajectories below and above it imply over- and under-confidence, respectively. The dark blue lines show that BayesANT produces well calibrated predictive probabilities on the test data, with a prediction accuracy equal to 85.2\% and 70.6\% and an average prediction probability of 0.82 and 0.70 in Scenarios 1 and 2, respectively. Results are based on adjusting initial probabilities with a temperature parameter $\rho$, trained on a hold-out dataset; see Materials and Methods.  
For the novel cases, the number of sequences predicted to belong to a ``new'' leaf in Scenario 1 is 958, while their true number is 884. Of these, 77.9\% are correctly recognized as novel, and 31.1\% are  effectively correct up to the \textit{Species} level, with average probability equal to 0.44 as depicted by the orange line. This implies that, while the exact ``new'' leaf in the taxonomy is generally difficult to retrieve, BayesANT recognizes fairly well the potential novelty of the taxon of a sequence.
Verification of the correctness of the novel branch requires further investigation - for example, by morphological assessment and/or more comprehensive sequencing of new samples collected at the same geographic location.  Similar results are obtained in Scenario 2.  While accuracy is lower due to a higher number of sequences with unobserved taxa, the predicted novel leaves are $2,736$ against $2,672$ truly ``new''. Here, $93.8\%$ are recognized novel, and $33.7\%$ are fully correct.


\begin{figure}[tbph]
    \centering
    \includegraphics[width =0.6\textwidth]{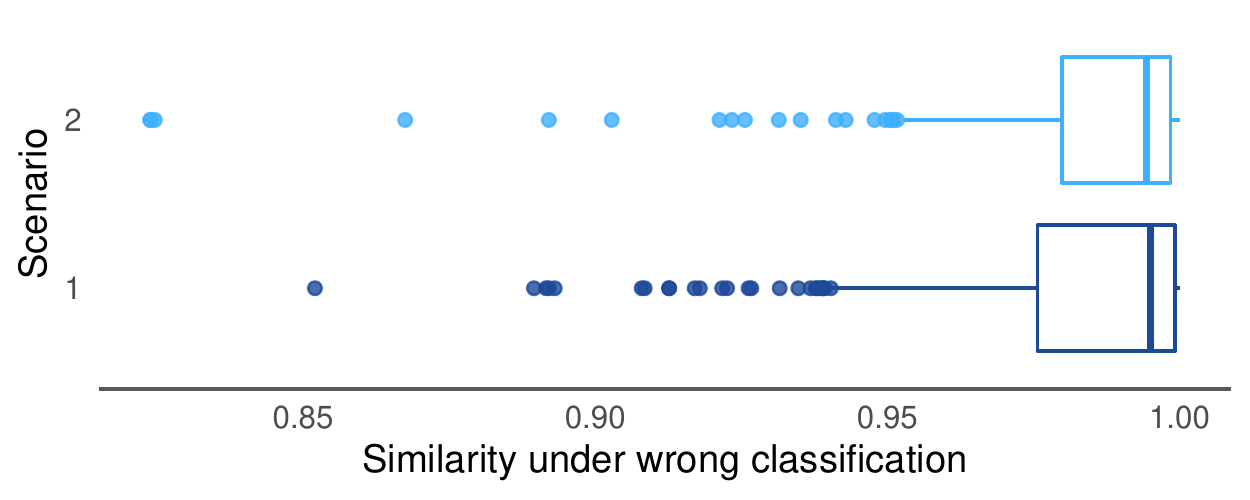}
    \caption{\small{Average DNA similarity between the test query sequences and the predicted taxa when BayesANT incorreclty predicts a taxon observed in training.}}
    \label{fig:wrong_similarity}
\end{figure}

In considering these taxonomic classification results, it is important to keep in mind the limited information provided by the available nucleotide sequencing in the COI gene.  This information can be insufficient to accurately assign certain query sequences to the correct taxon. To investigate whether this lack of information is a primary cause when BayesANT produces incorrect classifications, we measured the average similarity between the query test sequence and the sequences in the training which are annotated with the predicted taxa when the classification is wrong.  Figure~\ref{fig:wrong_similarity} shows the distribution of these average similarities.  Indeed, these are generally high, with an average of 
 0.983 under both Scenarios. This implies that misclassification tends to be due to insufficient information to distinguish between the true taxon and an incorrect taxon that is extremely close in the COI region to the query sequence.  The ability of BayesANT to provide accurate predictive probabilities is a major advantage in this limited information setting.

\subsection{Prediction accuracies}

As a last step in our analysis, we benchmark the performance of BayesANT on the FinBOL library against several alternatives in terms of accuracy. Table~\ref{tab:Accuracies_FinBOL} reports the results under both Scenarios. \textsc{m-1} refers to the single location multinomial kernel we adopted in our analysis above. While this is our method of reference due to its simplicity and flexibility, it treats loci as independent. Dependence can be introduced by adopting a 2-mer location kernel, \textsc{m-2},  where the support of the multinomial is in $\{\textrm{AA, AC, AT,} \ldots, \textrm{TT}\}$ and 2-mers  are overlapping. See the Materials and Methods section for details. To assess the advantage of adopting a Pitman--Yor prior over the taxonomic tree, we also compare with an analysis that lets $\alpha_\ell=\sigma_\ell=0$ at every level $
\ell$.  This does not allow new species, and the prior 
is the proportion with which each taxon appears in the library at every rank. These methods are labelled as $\textsc{m-1, no new}$ and $\textsc{m-2, no new}$ in the Table. Finally, we benchmark all these alternatives against the popular RDP classifier \citep[][version 2.13, 2020]{Wang_2007}. 

We first notice that no method is uniformly better than the others. Indeed, performances in Scenario 1 are approximately similar both in terms of prediction probabilities and accuracy. Minor differences are found at the \textit{Species} level, where the inclusion of novel taxa leads to higher accuracy for both \textsc{m-1} and \textsc{m-2}. Under novel taxa, all other methods are necessarily wrong. Moreover, the algorithms show a similar behavior in Scenario 2, which features a much higher proportion of unobserved taxa in training, with the exception of the \textit{Species} level. Here, BayesANT shows its advantage, as it attains a prediction that is 10\% more accurate than the RDP classifier. When we restrict to the \textit{Species} observed in training, however, model \textsc{m}-1 shows an accuracy of $93.1\%$ in Scenario 1 and of $93.7\%$ in Scenario 2, while RDP shows $95.3\%$ and $95.9\%$, respectively. The better performance of RDP under observed \textit{Species} is a consequence of the inclusion of taxonomic novelty, since extending the set of predictable taxa lowers their prior probability. Thus, BayesANT pays a price in terms of accuracy under observed taxa in favor of a much higher gain overall. Indeed, if we neglect novelty in BayesANT, the accuracies of \textsc{m-1, no new} on the observed \textit{Species} are $95.5\%$ and $96.7\%$.  

\begin{table}[tb]
\setlength{\tabcolsep}{4pt}
\begin{adjustbox}{max width=1\textwidth, center}
\begin{tabular}{lccccccc|ccccccc}
& \multicolumn{7}{c}{\textsc{scenario 1 - pure random split}} &  \multicolumn{7}{c}{\textsc{scenario 2 - stratified random split}}\\
\midrule
\textsc{model} &
\textsc{class} & \textsc{order} & \textsc{family} & \textsc{subfamily} & \textsc{tribe} & \textsc{genus} & \textsc{species} & 
\textsc{class} &\textsc{order} & \textsc{family} & \textsc{subfamily} & \textsc{tribe} & \textsc{genus} & \textsc{species} \\
\midrule
\textsc{m}-1 & \underline{100.0} & \underline{99.9} & \underline{98.6} & \underline{97.5} & 96.0 & 92.1 & 85.2 & \underline{99.8} & 97.0 & \underline{82.6} & \underline{80.8} & \underline{79.7} & 75.3 & \underline{70.6}\\
 & (1) & (1) & (.98) & (.96) & (.94) & (.91) & (.82) & (.99) & (.97) & (.87) & (.83) & (.8) & (.77) & (.7)\\
\textsc{m}-2 & \underline{100.0} & \underline{99.9} & 98.4 & 97.2 & 95.8 & 92.4 & \underline{85.4} & \underline{99.8} & \underline{97.1} & 82.1 & 80.3 & 79.1 & \underline{75.7} & 69.8\\
 & (1) & (1) & (.98) & (.97) & (.95) & (.93) & (.86) & (.98) & (.96) & (.88) & (.84) & (.81) & (.8) & (.74)\\
\textsc{m-1}, \textsc{no new} & \underline{100.0} & 99.5 & 98.0 & 97.2 & \underline{96.7} & \underline{94.3} & 83.3 & 98.7 & 91.8 & 75.8 & 75.3 & 74.6 & 72.1 & 59.4\\
 & (1) & (1) & (.99) & (.98) & (.98) & (.98) & (.92) & (1) & (.98) & (.91) & (.89) & (.89) & (.88) & (.78)\\
\textsc{m}-2,  \textsc{no new} & \underline{100.0} & 99.5 & 97.5 & 96.7 & 96.2 & 93.8 & 83.2 & 96.8 & 89.3 & 73.8 & 73.3 & 72.8 & 70.8 & 59.1\\
 & (1) & (1) & (.99) & (.98) & (.98) & (.98) & (.91) & (1) & (.98) & (.91) & (.89) & (.89) & (.88) & (.74)\\
\textsc{rdp} & \underline{100.0} & 99.6 & 97.9 & 97.1 & \underline{96.7} & 94.2 & 83.1 & 99.6 & 95.1 & 77.8 & 76.9 & 76.1 & 72.9 & 58.9\\
 & (1) & (.99) & (.97) & (.96) & (.95) & (.94) & (.92) & (.99) & (.92) & (.79) & (.78) & (.77) & (.75) & (.73)\\
\bottomrule
\end{tabular}
\end{adjustbox}
\caption{\small{Overall predictive performances of DNA barcoding algorithms on the FinBOL library under the two testing scenarios. Values report the \% of DNA sequences correctly labelled, while values in parenthesis denote the average prediction probabilities in the whole test set.  Underlined values indicate the best performances. } \label{tab:Accuracies_FinBOL}}
\end{table}

\section{Discussion}

This article has proposed a new probabilistic taxonomic classifier, BayesANT, which has the key property of allowing one to probabilistically build on an existing taxonomic library.  This is motivated by the fact that existing libraries for arthropods are incomplete, containing reference DNA sequences for a subset of the nodes of the taxonomic tree.  Some nodes lacking reference sequences correspond to branches of the tree not yet known to science. 
For example, it is estimated that approximately 1.5 million, 5.5 million, and 7 million species of beetles, insects, and terrestrial arthropods, respectively,  are either awaiting a proper description or are simply undiscovered \citep{Stork_2018}, with estimates varying every year. BayesANT uses species sampling priors \citep{Pitman1996} to allow for discovery of previously unknown branches of the tree, while characterizing uncertainty in all aspects of taxonomic classification including discovery of new species.  

Given that taxonomic classification in ecology studies typically relies on sequencing of a relatively short region of the genome, there is necessarily substantial uncertainty in classification \citep{Pentinsaari_2020}. For example, different species often have indistinguishable nucleotide sequences in the region being sequenced, so that it becomes impossible to reliably distinguish sequences from such species relying on DNA metabarcoding alone without supplemental morphology data. Hence, it is crucial that taxonomic classifiers provide a realistic measure of uncertainty in classification \citep{Somervuo_2016}.  Probabilistic forecasts providing accurate characterizations of predictive uncertainty are said to be well calibrated.  BayesANT
guarantees well calibrated predictions through a cross validation approach.

BayesANT builds on a popular species sampling prior known as the Pitman-Yor process \citep{Perman_1992}.  The classical Pitman-Yor process does not take into account the taxonomic tree structure, and instead treats all species as exchangeable. However, by using a Pitman-Yor at each level of the tree, with different parameters for each taxonomic rank, we obtain a highly flexible generative probabilistic process that can predict the probability of a new query sequence belonging to a different taxa at each level of the tree.  By estimating the Pitman-Yor parameters based on the training data, we allow the process to adapt to existing knowledge about the level of diversity at each taxonomic rank.

The modeling choices made in building BayesANT reflect a balance between flexibility and pragmatism in developing an efficient off-the-shelf algorithm that can easily handle classification of millions of sequences.  This is needed in our motivating applications to biodiversity monitoring studies that routinely collect and metabarcode samples from many different sites and multiple time points for each site.  In future research, it may be helpful to consider other modeling choices, which modify the Pitman-Yor structure  and/or choices of kernels considered here.
For example, instead of the simple multinomial kernels, it may be useful to explore similarity and latent variable-based likelihoods, for example using the projected $\kappa$-mer decomposition of a sequence into a lower dimensional feature space.

Taxonomic novelty due to missing branches in the reference libraries is  discussed in the literature \citep{Lan_2012,Somuervuo_2017}. Interpretation of the detected ``new'', however, is fairly delicate and context dependent, and requires further investigation on the sequenced DNA. It is common in practice to ignore the query sequences that show classification confidence below a certain threshold. Such an approach can be effective in removing reads suspected of sequencing error. However, it is unclear whether a low classification probability to existing taxa implies the sequence is likely from a novel taxa.  BayesANT provides a probabilistic framework for avoiding such arbitrary thresholds, instead characterizing uncertainty in all aspects of classification including to new branches of the tree.


\section{Materials and methods}

\bayesant evaluates the probabilities that a given DNA query sequence belongs to each of the nodes of the observed taxonomy. These probabilities are derived via Bayes rule. Let $V_i$ be the taxon of the $i$th sequence, and $\bX_i$ the associated nucleotide sequence in the COI gene. The probability that the query belongs to node $v$ is
\begin{equation}\label{eq:BayesRule}
    \text{pr}(V_i = v\mid \bX_i) \, \propto \, \pi_v p(\bX_i \mid V_i=v),
\end{equation}
where $\pi_v = \text{pr}(V_i = v)$ is the prior probability of taxon $v$ and $p(\bX_i \mid V_i=v)$ is the distribution of the DNA sequence conditioned on $v$ being the true label. In what follows, we carefully specify how each component is determined.  

\subsection{Notation and taxonomic structure} 
\begin{figure}[t]
\centering
\includegraphics[width=\textwidth]{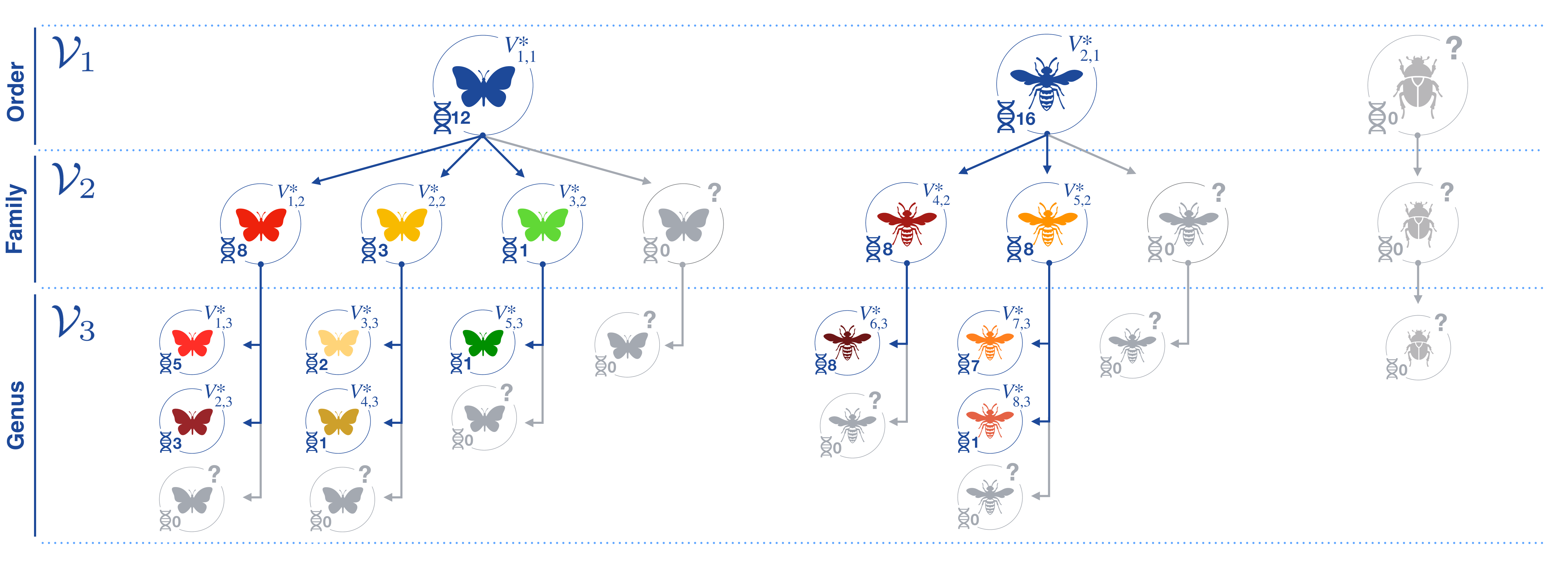}
\caption{\small{Example of a three-level taxonomic library under our model. On the bottom-left corner of every node we report the number of DNA sequences linked to it.  The total sample size of this example is $n =28$. Circles in blue indicate nodes linked to leaves with observed DNA sequences, while circles in gray show all the possible missing or undiscovered branches, labelled with a question mark on the top-right corner. Variation in insect color along each branch and across branches indicate DNA and morphological similarities and differences, respectively.}}\label{fig:Hierarchical_Taxonomy}
\end{figure}

A taxonomic library can be represented as a tree with branches of length $L\geq 2$, where DNA sequences are uniquely associated with one leaf.  We denote such a library as $\mathcal{D}_n = (\mathbf{V}_i, \mathbf{X}_i)_{i=1}^n$, where $n$ is the number of sequences,  $\bV_i = (V_{i, \ell})_{\ell=1}^L$ indicates the taxonomic labels of the $i$th sequence and $\bX_i$ is a representation of the associated DNA. Figure~\ref{fig:Hierarchical_Taxonomy} displays an example of a taxonomic tree where sequences are classified into \textit{Order}, \textit{Family} and  \textit{Genus}. Blue circles indicate nodes associated to at least one DNA sequence, while undiscovered branches are colored in grey. The labels at a given level $\ell$, namely $V_{1, \ell}, \ldots, V_{n, \ell}$, take values in the space $\mathcal{V}_\ell$ of distinct taxa. Given their discrete nature, multiple $V_{i, \ell}$ can be associated with the same taxon. These realizations, which we denote as $V_{1, \ell}^*, V_{2, \ell}^*, \ldots$ and so on, are the nodes in our hierarchical taxonomy. For example, the 28 sequences in Figure~\ref{fig:Hierarchical_Taxonomy} identify two taxa at the \textit{Order} level: one that has a butterfly-type morphological trait, $V_{1,1}^*$, and one with a bee-type trait, $V_{2,1}^*$. The beetle-shaped insect node instead represents a potential \textit{Order} yet to be discovered. 

Due to the tree structure of the taxonomy, 
each generic node $v_\ell$ at level $\ell$ has a unique parent at level $\ell -1$, denoted as $\textup{pa}(v_\ell)$. In Figure~\ref{fig:Hierarchical_Taxonomy}, for instance, $\pa(V_{1,2}^*) = V_{1,1}^*$ and  $\pa(V_{1,3}^*) = \pa(V_{2,3}^*) =  V_{1,2}^*$. For coherence, assume that the tree is rooted, namely $\pa(v_1) = v_0$ for any $v_1 \in  \mathcal{V}_1$. Each node in the tree is linked to multiple taxa at lower ranks. Let $\rho_n(v_\ell)$ be the set of observed nodes $v_{\ell+1}$ for which $\pa(v_{\ell+1}) = v_{\ell}$ when $n$ sequences are observed, $K_n(v_\ell) = |\rho_n(v_\ell)|$ be its cardinality and $N_n(v_\ell)$ be the number of DNA sequences belonging to $v_\ell$. In Figure~\ref{fig:Hierarchical_Taxonomy}, $\rho_n(V_{1,2}^*) = \{V_{1,3}^*, V_{2,3}^*\}$  and $K_n(V_{1,2}^*) = 2$, while $\rho_n(v_0) =\{V_{1,1}^*, V_{2,1}^*\}$  and $K_n(v_0) = 2$ for the \textit{Order} level. Finally, the size of a node in our representation is determined as a sum of the number of sequences stored in all leaves connected to it. For example, $N_n(V_{1,2}^*) = 8$, and $N_n(V_{1,1}^*) = 12$. The quantities $\pa(\cdot)$, $\rho_n(\cdot)$, $K_n(\cdot)$ and $N_n(\cdot)$ are the key ingredients upon which we build our taxonomic prior $\pi_v$ in equation~\eqref{eq:BayesRule}.
 
\subsection{Taxonomic prior}
The first step in our analysis consists of specifying a flexible prior for the frequencies of occurrence of different types of organisms at each taxonomic rank 
$\ell$, including organisms of ``new'' types.
In particular, we incorporate the 
 Pitman--Yor process allocation probabilities in equations~\eqref{eq:PY_probOld} and~\eqref{eq:PY_probNew} into the tree structure. Let $\alpha_\ell$ and $\sigma_\ell$ denote the allocation parameters for level $\ell$, with $\alpha_\ell > -\sigma_\ell$ and $\sigma_\ell \in [0,1)$. Write $\mathbf{V}_{\cdot, \ell}^{(n)} = (V_{i, \ell})_{i=1}^n$ as the sequence of taxonomic labels observed at level $\ell$. Then, the taxon of sequence $n+1$ at level $\ell$, conditioned on it being allocated to node $v_{\ell-1}$ at level $\ell - 1$, has probabilities 
\begin{equation}\label{eq:PY_probTaxaOld}
    \text{pr}(V_{n+1, \ell} = V_{j,\ell}^* \mid V_{n+1, \ell-1} = v_{\ell-1}, \mathbf{V}_{\cdot, \ell}^{(n)}) = \frac{N_n( V_{j,\ell}^* )\! - \! \sigma_\ell}{\alpha_\ell\! + \!N_n(v_{\ell-1})},
\end{equation}
if the node $V_{j, \ell}^*$ is such that $\pa( V_{j,\ell}^*) = v_{\ell - 1}$, and 

\begin{equation}\label{eq:PY_probTaxaNew}
    \!\!\text{pr}(V_{n+1, \ell}\! =\!\text{``new''}\! \mid \! V_{n+1, \ell-1}\! =\! v_{\ell-1}, \mathbf{V}_{\cdot, \ell}^{(n)})\! =\! \frac{\alpha_\ell\! + \!\sigma_\ell K_n(v_{\ell-1})}{\alpha_\ell \!+ \!N_n(v_{\ell-1})},
\end{equation}
if the node is new and originates from $v_{\ell - 1}$. The structure of equations~\eqref{eq:PY_probTaxaOld} and  \eqref{eq:PY_probTaxaNew} is the same as the one in equations~\eqref{eq:PY_probOld} and \eqref{eq:PY_probNew}, with the only difference being that nodes at $\ell$ are generated from their parent-specific process. The level-specific parameters $\alpha_\ell$ and $\sigma_\ell$ are important in allowing diversity to vary with taxonomic rank.  These parameters will be estimated based on the data as we detail in the Supplementary material.

\subsection{DNA sequence likelihood}
The second step to build the predictive rule in equation~\eqref{eq:BayesRule} is to specify a distribution for the DNA sequences. We do this by adopting a kernel-based approach that flexibly accommodates different DNA representations. 

As depicted in Figure~\ref{fig:Hierarchical_Taxonomy}, a query sequence $\bX_i$ is uniquely associated with one leaf of the taxonomic tree. Recalling that $V_{i, L}$ is the taxon of the $i$th sequence at the lowest level $L$, we let 
\begin{equation}\label{eq:Kernel}
    (\bX_i \mid V_{i, L} = v_L, \btheta_{v_L}) \stackrel{\text{iid}}\sim \mathcal{K}(\bX_i; \btheta_{v_L}),
\end{equation}
where $v_L \in \mathcal{V}_L$ is a generic leaf, 
$\mathcal{K}(\bX_i;\btheta)$ is a kernel depending on parameters $\btheta$ representing the likelihood of sequence data $\bX_i$, and 
$\btheta_{v_L}$ is a collection of leaf-specific parameters.  We assume that all DNA sequences associated to leaf $v_L$ are independent and identically distributed as $\mathcal{K}(\cdot; \btheta_{v_L})$. Table~\ref{tab:Multinomial} provides three examples of multinomial-type kernels when sequences are aligned and when they are not. Here, \textit{alignment} implies that all the sequences are pre-processed to have the same length $p$ so that the nucleotides at each position $s=1, \ldots, p$  are meaningfully comparable. Then, $X_{i, s}$ is the nucleotide in the $s$th position of the $i$th query sequence, and $\theta_{v_L, s, g}$ is the probability that nucleotide $g \in \mathcal{N}_1 = \{\text{A, C, G, T}\}$ is seen at $s$ for taxon $v_L$. Assuming independence across locations $s$ as a simplifying assumption to improve computational efficiency in constructing a probabilistic classifier, the resulting kernel is a product of 
multinomials with location-specific parameters. 

\begin{table}[tb]
\centering
\setlength{\tabcolsep}{4pt}
\begin{tabular}{lccc}
\textsc{sequences} & \textsc{method}  &\textsc{kernel} & \textsc{prior} $\btheta_{v_L}$ \\
\toprule
Not aligned & $\kappa$-mers & $\prod_{g \in \mathcal{N}_\kappa} \theta_{v_L \!, g}^{n_{i\!,g}}$ & $\textsc{dir}(\boldsymbol{\xi}_{v_L})$ \\ [1.05ex] 
Aligned & Product & $\prod_{s=1}^p \! \prod_{g \in \mathcal{N}_1} \theta_{v_L,s,g}^{\mathbf{1}\{X_{i\!,s} = g\} }$ &  $\prod_s \!\textsc{dir}(\boldsymbol{\xi}_{v_L\!,s})$\\
Aligned & $\kappa$-Product & $\prod_{s=1}^p \! \prod_{g \in \mathcal{N}_\kappa} \theta_{v_L,s,g}^{\mathbf{1}\{X_{i\!,s} = g\} }$ &  $\prod_s \!\textsc{dir}(\boldsymbol{\xi}_{v_L\!,s})$\\
\bottomrule
\end{tabular}
\caption{\small{Examples of multinomial kernels for the DNA sequences. $\mathcal{N}_\kappa$ is the set of all $\kappa$-mers on which the sequence is decomposed. In the aligned case, this is a set of monomers $\mathcal{N}_1 = \{\text{A,C,G,T}\}$. The quantity $\mathbf{1}\{X_{i,s} = g\}$ is an indicator equal to one if $X_{i,s} = g$ and zero otherwise.} \label{tab:Multinomial}}
\end{table}

When sequences are \textit{not} aligned, each has its own length $p_i$. A viable option is to use a \kmer decomposition. 
This amounts to counting the number of times all possible $4^\kappa$ substrings of length $\kappa$ appear within the sequence.  We denote as $\mathcal{N}_\kappa$ the set of all \kmers of length $\kappa$. For instance, 3-mers live in $\mathcal{N}_3 = \{\text{AAA, ACG, AGT} \ldots \}$, with a total of $4^3 = 64$ substrings. In Table~\ref{tab:Multinomial}, $n_{i, g} = \sum_{s=1}^{t_i} \mathbf{1}\{X_{i,s}=g\}$ denotes the number of times a \kmer $g\in \mathcal{N}_\kappa$ appears in the $i$th sequence, with $t_i=p_i-\kappa+1$ being the total number of \kmers observed when the length is $p_i$. We model these counts as the output of a multinomial distribution, where $\theta_{v_L, g}$ is the probability of \kmer g at taxon $v_L$. 

The choice of kernel depends on the application and on the data. Insect DNA sequences can be easily aligned via hidden Markov models \citep{Eddy_1995}. This is not true for fungal sequences, which are substantially more diverse and hence difficult if not impossible to align. Irrespective of the structure of the data, our proposed multinomial kernels have the advantage of simplicity in computation, with the posterior distribution for $\theta_{vL}$ obtained in analytic form by adopting conjugate Dirichlet priors as in Table~\ref{tab:Multinomial}.  Computational efficiency is a critical issue in classifying very large numbers of sequences, making it intractable to consider elaborate likelihoods derived from realistic generative models of nucleotide sequences. 

\subsection{Prediction rule}
The prior on the tree and the DNA sequence likelihood defined so far allow us to predict the set of labels $\bV_{n+1}$ for the query sequence $\bX_{n+1}$. BayesANT does this in \textit{bottom-up} and \textit{top-down} steps. In the \textit{bottom-up} step, we use equations~\eqref{eq:PY_probTaxaOld}, \eqref{eq:PY_probTaxaNew} and \eqref{eq:Kernel} to determine the posterior probability that $\bX_{n+1}$ belongs to \textit{any} leaf in the tree. These include both the observed and the new taxa at the lowest level, as illustrated in Figure~\ref{fig:Hierarchical_Taxonomy}\footnote{Under the assumption that a new node at level $\ell$ automatically creates a new node at all levels $\ell + 1, \ldots, L$ below, the total number of potentially unobserved leaves is equal to the number of nodes up to $L-1$ plus 1}. Then, probabilities of higher nodes are computed aggregating upward. In the \textit{top-down} step, instead, BayesANT predicts a branch by iteratively choosing the child node with the highest probability at each level, starting from the root.  

Let $\pi_{n+1}(v_L)$ be the prior probability of leaf $v_L$ after having observed $\mathcal{D}_{n}$. This is equal to the product of the prior conditional probabilities in equations~\eqref{eq:PY_probTaxaOld} and \eqref{eq:PY_probTaxaNew} of all nodes in the branch $v_0$-$v_L$, which is 
\begin{equation}\label{eq:PriorProb}
     \pi_{n+1}(v_L)  =  \text{pr}(V_{n+1, 1} = v_1 \mid \mathbf{V}_{\cdot, 1}^{(n)}) \prod_{\ell=2}^{L} \text{pr}(V_{n+1, \ell}=v_{\ell}\mid V_{n+1, \ell-1}=v_{\ell-1}, \mathbf{V}_{\cdot, \ell}^{(n)}).
\end{equation}
Equation~\eqref{eq:PriorProb} is the equivalent of the  prior taxon probability in equation~\eqref{eq:BayesRule}. Notice that if $v_\ell = \text{``new''}$ at some $\ell$, the conditional probabilities at lower nodes are equal to 1. But then, the probability that $V_{n+1, L}$ is associated to taxon $v_L$ \textit{conditioned on the DNA sequence} $\bX_{n+1}$ is

\begin{equation}\label{eq:LeafProb}
     p_{n+1}(v_L)  = \text{pr}(V_{n+1, L} = v_L \mid \bX_{n+1}, \mathcal{D}_n)  \propto \pi_{n+1}(v_L) \int \K(\bX_{n+1}; \btheta_{v_L})p(\btheta_{v_L} \! \mid\!  \mathcal{D}_n)\text{d}\btheta_{v_L}.
\end{equation}
The integral in equation~\eqref{eq:LeafProb} is the posterior predictive distribution of DNA sequence $\bX_{n+1}$ with respect to the posterior of $\btheta_{v_L}$. When $v_\ell = \text{``new''}$, this posterior is equal to the prior, i.e. $p(\btheta_{v_L} \! \mid\!  \mathcal{D}_n)= p(\btheta_{v_L})$, as no sequence for $v_L$ is observed. The advantage of the models in Table~\ref{tab:Multinomial} is that both the prior and the posterior predictive distribution have simple and easy-to-compute analytic forms. 

Once equation~\eqref{eq:LeafProb} has been evaluated for all leaves, the probabilities of higher nodes in the taxonomy can be easily derived via upward aggregation. Then, we predict the taxa by starting from the root of the tree and recursively selecting the child node with the highest probability. Specifically, the predicted sequence of taxa
$(v_\ell^*)_{\ell=1}^L$ for the DNA sequence at $n+1$ satisfies
\begin{equation}\label{eq:PredictionRule}
    v^*_\ell = {\arg\max}_{v_\ell \in \rho_n(v^*_{\ell-1})} \sum_{v_L \in \mathcal{L}_n(v_\ell)} p_{n+1}(v_L),
\end{equation}
where $\mathcal{L}_n(v_\ell)$ is the set of all leaves linked to node $v_\ell$ in a library of $n$ DNA sequences. 

\subsection{Hyperparameter tuning} 
The hyperparameters $\boldsymbol{\xi}_v$ of the multinomial kernel play a fundamental role in novel species recognition. As detailed above, when $v_L = \text{``new''}$, then equation~\eqref{eq:LeafProb} is a prior predictive probability, since no sequence is observed for $v_L$ and  thus $p(\btheta_{v_L} \! \mid\!  \mathcal{D}_n)= p(\btheta_{v_L})$. In such cases, prior hyperparameters should contain information regarding the taxonomic branch and level where novelty appears.  Uniform priors may be unreasonably vague, leading to under-estimation of the prior predictive probability of novel taxa relative to the true proportion. Thus, we tune each $\boldsymbol{\xi}_{v_L}$ as follows. Consider a taxon $v_{L-1}$ at level $L-1$. If $v_{L-1}$ is not ``new'', the hyperparameters $\boldsymbol{\xi}_{v_L}$ of all the leaves $v_L \in \mathcal{L}_n(v_{L-1})$ linked to it - including the new one - are all equal, and they are obtained via method of the moments from the DNA sequences $\bX_{i}$ with $V_{i,L-1} = v_{L-1}$. If instead $v_{L-1}$ is a ``new'' node and the last not novel node in its branch is $v_\ell$ at level $\ell \leq L-1$, the method of the moments is applied on the set of sequences $\bX_{i}$ such that $V_{i,\ell} = v_\ell$. This ensures borrowing of information between the branches when the novelty appears at higher levels in the taxonomy.
For mathematical details on the method of the moments applied to the multinomial kernels of Table~\ref{tab:Multinomial}, see the Supplementary material.


\subsection{Calibration of prediction probabilities}
Misspecification of a Bayesian model, due to inaccuracies in the prior and/or likelihood function, may lead to predictive probabilities that are not sufficiently well calibrated to accurately capture predictive uncertainties \citep{Grunwald_Ommen_2017, Miller_Dunson_2019}.  Given the complexity of the true data-generating likelihood underlying DNA barcoding data, and the necessity of using a simple likelihood for computational tractability, some degree of misspecification is inevitable. To adjust the predictive probabilities used in equation~\eqref{eq:PredictionRule} for misspecification, we apply a simple re-calibration approach. 
In particular, we post-process the prediction probabilities in  equation~\eqref{eq:LeafProb} by exponentiating them by a coefficient $\rho  \in (0,1]$ and later renormalizing. Then, the new  probabilities for the $(n+1)$st sequence are 
\begin{equation}\label{eq:CalibratedProbs}
    \tilde{p}_{n+1}(v_L) = \frac{p_{n+1}(v_L)^\rho}{\sum_{v \in \mathcal{V}_L} p_{n+1}(v)^\rho},
\end{equation}
and can be used in place of $p_{n+1}(v_L)$ in equation~\eqref{eq:PredictionRule}. Such a strategy does not alter the ranking of the original probabilities since the transformation is monotonic. Moreover, if $p_{n+1}(v_L)=1$, then also $\tilde{p}_{n+1}(v_L)=1$. This implies that we do not substantially alter the prediction whenever the BayesANT is certain about a taxon. Choices for $\rho$ can be adopted via cross validation on an hold-out subset of the training library following strategies such as the ones described in \citep{Guo_2017}. Specifically, prediction probabilities are calibrated if the average probability for the predicted nodes is equal to the classification accuracy \citep{Somervuo_2016}. For example, if 90\% of the sequences are correctly classified, ideally the average classification probability is approximately 0.9. An average value of 0.5 and of 0.99, instead, means that the algorithm is too conservative when right and too confident when wrong, respectively. We select $\rho = 0.1$ under both Scenarios of the FinBOL application. In general, our experience suggests that $\rho \approx 0.1$ works well in practice with other libraries as well.

\subsection*{Software availability}
BayesANT is available as an open-source \texttt{R} package at \url{https://alessandrozito.github.io/BayesANT/vignette.html}. 

\subsection*{Data availability}
We use the full COI version of the library available at \url{https://github.com/psomervuo/FinPROTAX}. We restrict to sequences with fully specified taxonomy up to the \textit{Species} level. 

\subsection*{Acknowledgements}
This project has received funding from the European Research Council under the European Union’s Horizon 2020 research and innovation programme (grant agreement No 856506). The authors would like to express their gratitude to Otso Ovaskainen, Panu Somervuo, Jesse Harrison, Markus Koskela, Tomas Roslin, Bianca Dumitrascu and Jennifer Kampe for their precious suggestions, to Elena Domenichini for graphical advice and to Carolyn Quarterman for the support on the writing.

\bibliographystyle{chicago}
\bibliography{pnas-sample}

\newpage

\appendix

\section*{Supplementary Material}
The following document contains mathematical and theoretical details for the BayesANT algorithm - BAYESiAn Nonparametric Taxonomic classifier - described in the main paper. Emphasis is on the explicit formulas for the taxonomic annotation probabilities and the associated estimation method for the model parameters. 

\section{Prior taxonomic probabilities}
In this section we provide details on the prior probabilities over the nodes in the taxonomic tree, and describe the estimation procedure for the associated hyperparameters. 

\subsection{The Pitman--Yor process and the exchangeable partition probability function}
BayesANT models taxonomic novelty via Pitman--Yor process priors \citep{Perman_1992}. As already detailed in the main paper,  the process works as follows. Let $V_1,\ldots, V_n$ be a sequence of taxon assignments for the DNA sequences in the training library,  comprising of a total of $K_n = k$ distinct labels denoted as $V_1^*, \ldots, V_k^*$ and appearing with frequencies $n_1, \ldots, n_k$. Then, the taxon of the $(n+1)$st observation is determined via the following allocation scheme:

\begin{equation}\label{eq:PY_basic}
(V_{n+1} \mid V_1, \ldots, V_n) = 
\begin{cases}
        V_j^*, & \text{with prob.} \quad (n_j - \sigma)/(\alpha + n), \ j = 1, \ldots, k,\\
        \text{``new''}, & \text{with prob.} \quad (\alpha + \sigma k)/(\alpha + n),
\end{cases}
\end{equation}
where $\sigma \in [0,1)$ is a discount parameter governing the tail of the process and $\alpha > -\sigma$ is a precision parameter. High values for  $\alpha$ and $\sigma$ lead to a high number of distinct labels $K_n$. Moreover,  high values for $n_j$ lead to a high probability that taxon $V_j^*$ will be observed in the future. See Figure 1 in the main paper for a practical illustration. Estimation of both parameters can be performed via an empirical Bayes procedure \citep{Favaro_lijoi_mena_pruenster_2009} through maximization of the quantity known as \textit{exchangeable partition probabilitity function} \citep[EPPF,][]{Pitman1996}. Let $N_j$ denote the random variable corresponding to the frequency of appearance of taxon $V_j^*$, with $n_j$ the realization of this random variable for $K_n = k$. The EPPF is defined as
\begin{equation}\label{eq:eppf}
\text{pr}(K_n = k, N_1 = n_1, \ldots, N_k = n_k)= \frac{\prod_{i = 1}^{k-1}(\alpha + i\sigma)}{(\alpha + 1)_{n-1}}\prod_{j=1}^k (1-\sigma)_{n_j -1},
\end{equation}
where $(x)_a = \Gamma(x+a)/\Gamma(x)$ is the Pochhammer factorial and $\Gamma(x)$ is the gamma function. The quantity in equation~\eqref{eq:eppf} can be interpreted as a likelihood function arising from the process in equation~\eqref{eq:PY_basic}. Then, one can simply apply maximum likelihood estimation as 
$$
(\hat{\alpha}, \hat{\sigma}) = {\arg\max}_{\alpha, \sigma} \Bigg\{\frac{\prod_{i = 1}^{k-1}(\alpha + i\sigma)}{(\alpha + 1)_{n-1}}\prod_{j=1}^k (1-\sigma)_{n_j -1} \Bigg\}, \quad \sigma\in[0,1), \alpha > -\sigma.
$$
In what follows, we apply this procedure to estimate the parameters $\alpha_\ell$ and $\sigma_\ell$ for all levels $\ell = 1, \ldots, L$ in the taxonomic tree.

\subsection{Level-specific Pitman--Yor priors} Consider a taxonomic library $\mathcal{D}_n = (\mathbf{V}_i,\mathbf{X}_i )_{i=1}^n$ of size $n$ and of $L\geq 2$ levels, where $\mathbf{V}_i = (V_{i, \ell})_{\ell =1}^L$ are the taxonomic annotations for DNA sequence $\mathbf{X}_i$. Following the notation in the main paper, we let  $V_{j, \ell}^*$ be the $j$th taxon and $\mathbf{V}_{\cdot, \ell}^{(n)} = (V_{i, \ell})_{i =1}^n$  be the sequence of taxa observed for level $\ell$. To construct the taxonomic tree, we introduce the following quantities. For a generic taxon $v_\ell$ at level $\ell$, we define $\textrm{pa}(v_\ell)$ as the unique parent node of $v_\ell$ at level $\ell -1$ and $\rho_n(v_\ell)$ as the set of nodes $v_{\ell+1}$ at level $\ell+1$ such that $\textrm{pa}(v_{\ell+1}) = v_\ell$. We also let $K_n(v_\ell) = |\rho_n(v_\ell)|$ be the number of nodes linked to $v_\ell$ at level $\ell + 1$ and $N_n(v_\ell)$ be the size of the taxon, namely the number of DNA sequences linked to $v_\ell$. Then, BayesANT follows the prediction scheme in equation~\eqref{eq:PY_basic} by letting 
\begin{equation}\label{eq:PY_level_l}
(V_{n+1, \ell} \mid V_{n+1, \ell-1} = v_{\ell-1}, \mathbf{V}_{\cdot, \ell}^{(n)}) = 
\begin{cases}
        V_{j, \ell}^*, & \text{with prob.} \quad \big(N_n(V_{j, \ell}^*) - \sigma_\ell \big)/\big(\alpha_\ell + N_n(v_{\ell-1})\big),\\
        \text{``new''}, & \text{with prob.} \quad \big(\alpha_\ell + \sigma_\ell K_n(v_{\ell-1})\big)/\big(\alpha_\ell + N_n(v_{\ell-1})\big),
\end{cases}
\end{equation}
for $j: \textrm{pa}(V_{j, \ell}^*) = v_{\ell -1}$, where $\sigma_\ell \in [0,1)$ and $\alpha_\ell > -\sigma_\ell$ are rank-specific parameters. Equation \eqref{eq:PY_level_l} holds independently for all the observed nodes $v_{\ell-1}$ at level $\ell - 1$. 
Specifically, we model all the separate sets of taxa $\rho_n(v_{\ell-1})$ at a given rank $\ell$ as realizations from independent Pitman--Yor processes. In
estimating parameters $\alpha_\ell,\sigma_\ell$, we borrow of information across branches. The level-specific EPPF is a product EPPFs, and the estimates for $\alpha_\ell$ and $\sigma_\ell$ are obtained as
\begin{equation}\label{eq:PY_max}
(\hat{\alpha}_\ell, \hat{\sigma}_\ell) = {\arg\max}_{\alpha_\ell, \sigma_\ell} \Bigg\{
\prod_{v\in \mathcal{V}_{\ell-1}}\frac{\prod_{i = 1}^{K_n(v)-1}(\alpha_\ell + i\sigma_\ell)}{(\alpha_\ell + 1)_{N_n(v)-1}}\prod_{v_\ell \in \rho_n(v)} (1-\sigma_\ell)_{N_n(v_\ell) -1} \Bigg\}, 
\end{equation}
for $\sigma_\ell\in[0,1), \alpha_\ell > -\sigma_\ell$, where $\mathcal{V}_{\ell-1}$ denotes the set of all taxonomic nodes $V_{j, \ell-1}^*$ observed in the library at level $\ell-1$. The maximization  in equation~\eqref{eq:PY_max} can easily be carried out with routine methods such as \texttt{nlminb} in \texttt{R}. 

\subsection{Leaf probabilities} After $\alpha_\ell$ and $\sigma_\ell$ have been estimated from the library $\mathcal{D}_n$, BayesANT specifies the prior probability for each leaf node $v_L \in \mathcal{V}_L$, including new ones, as a product of Pitman--Yor probabilities. For the $(n+1)$st sequence, this is equal to
\begin{equation}\label{eq:priorProb}
\begin{split}
       \pi_{n+1}(v_L) &=  \text{pr}(V_{n+1, L} 
       = v_L \mid \mathcal{D}_n) \\
       &= \text{pr}(V_{n+1, 1} = v_1 \mid \mathbf{V}_{\cdot, 1}^{(n)})\times \prod_{\ell=2}^{L} \text{pr}(V_{n+1, \ell}=v_{\ell}\mid V_{n+1, \ell-1}=v_{\ell-1}, \mathbf{V}_{\cdot, \ell}^{(n)}).
\end{split}
\end{equation}
To see this explicitly, consider the example of a $L=4$ level library of size $n$ and let $V_{1, 1}^* \rightarrow V_{1, 2}^* \rightarrow V_{1, 3}^*\rightarrow V_{1,4}^*$ be a branch. This is a fully observed branch, for which  $\text{pa}(V_{1, 2}^*)= V_{1, 1}^*$, $\text{pa}(V_{1, 3}^*)= V_{1, 2}^*$ and $\text{pa}(V_{1, 4}^*)= V_{1, 3}^*$. Then, the prior probability for the leaf node $V_{1,4}^*$ is

$$
\pi_{n+1}(V_{1,4}^*) = \underbrace{\frac{N_n(V_{1, 1}^*) - \sigma_1}{\alpha_1 + n}}_{\substack{\textrm{Prob. of choosing} \\ V_{1,1}^* \textrm{ at Level 1}}} \times
\underbrace{\frac{N_n(V_{1, 2}^*) - \sigma_2}{\alpha_2 + N_n(V_{1,1}^*)}}_{\substack{\textrm{Prob. of choosing} \\ V_{1,2}^* \textrm{ at Level 2}}} \times
\underbrace{\frac{N_n(V_{1, 3}^*) - \sigma_3}{\alpha_3 + N_n(V_{1,2}^*)}}_{\substack{\textrm{Prob. of choosing} \\ V_{1,3}^* \textrm{ at Level 3}}} \times
\underbrace{\frac{N_n(V_{1, 4}^*) - \sigma_4}{\alpha_4 + N_n(V_{1,3}^*)}}_{\substack{\textrm{Prob. of choosing} \\ V_{1,4}^* \textrm{ at Level 4}}}.
$$
Consider instead the path $V_{1, 1}^* \rightarrow V_{1, 2}^* \rightarrow\text{``new''} \rightarrow\text{``new''}$. This identifies a new branch at level $\ell =3$, which in turn creates a new leaf. We denote it as $v_{L}^{\text{new}}$. Then, its associated probability is 

$$
\pi_{n+1}(v_{L}^{\text{new}}) = \underbrace{\frac{N_n(V_{1, 1}^*) - \sigma_1}{\alpha_1 + n}}_{\substack{\textrm{Prob. of choosing} \\ V_{1,1}^* \textrm{ at Level 1}}} \times
\underbrace{\frac{N_n(V_{1, 2}^*) - \sigma_2}{\alpha_2 + N_n(V_{1,1}^*)}}_{\substack{\textrm{Prob. of choosing} \\ V_{1,2}^* \textrm{ at Level 2}}} \times
\underbrace{\frac{\alpha_3 + \sigma_3K_n(V_{1,2}^*)}{\alpha_3 + N_n(V_{1,2}^*)}}_{\substack{\textrm{Prob. of novelty} \\ \textrm{at Level 3}}} \times
\underbrace{1}_{\substack{\textrm{Prob. of novelty} \\ \textrm{at Level 4}}}.
$$
Here, the novelty probability of the Pitman--Yor process appears at level 3. At level 4 the probability is equal to one since the node is necessarily new and $K_n(v_L^{\text{new}}) = N_n(v_L^{\text{new}})=0$. In a similar fashion, the probability for the branch $\text{``new''}\rightarrow\text{``new''} \rightarrow\text{``new''}\rightarrow\text{``new''}$ is 
$$
\pi_{n+1}(v_{L}^{\text{new}}) = \underbrace{\frac{\alpha_1 + \sigma_1K_n(v_0)}{\alpha_1 + n}}_{\substack{\textrm{Prob. of novelty} \\ \textrm{at Level 1}}} \times
\underbrace{1}_{\substack{\textrm{Prob. of novelty} \\ \textrm{at Level 2}}} \times
\underbrace{1}_{\substack{\textrm{Prob. of novelty} \\ \textrm{at Level 3}}} \times
\underbrace{1}_{\substack{\textrm{Prob. of novelty} \\ \textrm{at Level 4}}},
$$
since the novelty is detected first at level $\ell =1$. Finally, notice that a branch such as $V_{1, 1}^* \rightarrow V_{1, 2}^* \rightarrow\text{``new''} \rightarrow V_{1, 4}^*$  is not allowed in our representation.  
Under such a formulation, we are able to specify all the prior probabilities for both the observed taxa and the new ones in a coherent way. 

\section{Posterior taxonomic probabilities}
BayesANT assumes DNA sequences $\mathbf{X}_i$ associated with leaf  
$v_L \in \mathcal{V}_L$ are distributed as 
$$
(\mathbf{X}_i \mid V_{i, L} = v_L, \btheta_{v_L}) \stackrel{\text{iid}}{\sim}\mathcal{K}(\bX_i; \btheta_{v_L}),
$$
where $\mathcal{K}(\cdot, \btheta_{v_L})$ is a distribution that depends on the leaf-specific vector of parameters $\btheta_{v_L}$.  In what follows, we provide mathematical details of the single location product-multinomial model we use in the main document. Adapting the details to accommodate alternative kernels is straightforward.

\subsection{Multinomial kernel}
Let $\mathbf{X}_i$, $i= 1,\ldots, n$, indicate a collection of \textit{aligned} DNA sequences of length $p$. The alignment of the sequences makes the individual loci comparable across taxa.
An example for $p=20$ loci is as follows:
\begin{table}[h]
\centering
\begin{tabular}{lcccccccccccccccccccc}
\textsc{locus} $s$ &1&2&3&4&5&6&7&8&9&10&11&12&13&14&15&16&17&18&19&20\\
\midrule
$\bX_{1}$: & A & C &C &T &C &G &G &A &A &A &T& T& T& G& G& A& A& T& C& A \\
$\bX_{2}$: &A&C&T&T&C&G&A&A&T&A&T&A&A&G&A&G&A&T&G&G \\
$\bX_{3}$: &A&T&T&C&C&G&T&A&G&G&T&T&T&G&A&G&T&T&G&A 
\end{tabular}
\end{table}

As each loci in each sequence corresponds to a nucleotide in $\mathcal{N}_1=\{\textrm{A, C, G, T}\}$, it is natural to use a multinomial likelihood, 
$$
(X_{i, s} \mid V_{i, L} = v, \btheta_{v, s}) \stackrel{\text{iid}}{\sim} \textsc{mult}(1; \theta_{v, s, \text{A}}, \theta_{v, s, \text{C}},\theta_{v, s, \text{G}}, \theta_{v, s, \text{T}}),
$$
where $\btheta_{v, s} = (\theta_{v, s, \text{A}}, \theta_{v, s, \text{C}},\theta_{v, s, \text{G}}, \theta_{v, s, \text{T}})^\top$ is a vector of probabilities summing up to 1, and $\theta_{v, s,g}$ is the probability of observing nucleotide $g$ in position $s$ for leaf $v$. To simplify the analysis and ease computation, we further assume that all locations $s$ are independent, leading to the following likelihood contribution for the $i$th sequence:  
\begin{equation}\label{eq:multinomial}
\mathcal{K}(\bX_i; \btheta_{v}) = \prod_{s=1}^p\prod_{g\in \mathcal{N}_1} \theta_{v, s,g}^{\mathbf{1}\{X_{i,s} = g\}}, 
\end{equation}
where $\mathbf{1}\{\cdot\}$ denotes the indicator function. As a conjugate prior for the nucleotide probabilities, we choose
 $$\btheta_{v,s} \sim \textsc{dirichlet}(\xi_{v,s,\text{A}}, \xi_{v,s,\text{C}}, \xi_{v,s,\text{G}}, \xi_{v,s,\text{T}}),$$
with $\bxi_{v,s} = (\xi_{v,s,\text{A}}, \xi_{v,s,\text{C}}, \xi_{v,s,\text{G}}, \xi_{v,s,\text{T}})^\top$ a vector of  hyperparameters. The posterior distribution for the nucleotide probabilities at locus $s$ under leaf $v$ conditional on the DNA sequences assigned to $v$ is then
$$
(\btheta_{v,s} \mid \mathcal{D}_n) \sim \textsc{dirichlet}(\xi_{v,s,\text{A}} + n_{v,s,\text{A}}, 
\xi_{v,s,\text{C}} + n_{v,s,\text{C}}, 
\xi_{v,s,\text{G}}+ n_{v,s,\text{G}}, 
\xi_{v,s,\text{T}} + n_{v,s,\text{T}}),
$$
where 
 $n_{v,s,g} = \sum_{i: V_{i, L} = v} \mathbf{1}\{X_{i,s} = g\}$ indicates the number of times nucleotide $g \in \mathcal{N}_1$ is recorded at locus $s$ for the DNA sequences linked to leaf $v$. The resulting posterior kernel for $\btheta_v$ is then a product of independent Dirichlet distributions, namely
 \begin{equation}\label{eq:posterior}
     p(\btheta_v \mid \mathcal{D}_n)  \propto \prod_{s=1}^p \prod_{g\in \mathcal{N}_1} \theta_{v, s,g}^{\xi_{v,s,g} + n_{v,s,g}}.
 \end{equation}
An equivalent representation can be obtained for the 2-mer location multinomial kernel detailed in the main paper, but with the support of the multinomial being all pairs of nucleotides instead of singletons.
When sequences are not aligned, 
 the posterior in equation~\eqref{eq:posterior}
 is modified to remove the product 
 from $s=1, \ldots, p$ and substitute 
 $n_{v,s,g}$ with $n_{v,g}$, which is the number of times $\kappa$-mer $g$ is recorded in the sequence.

\subsection{Prior and posterior predictive distribution}
Once the parameters for the posterior distribution in equation~\eqref{eq:posterior} have been computed for each leaf node $v_L$, BayesANT determines the prediction probabilities $p_{n+1}(v_L) = \text{pr}(V_{n+1, L} = v_L \mid \bX_{n+1}, \mathcal{D}_n)$ as
\begin{equation}\label{eq:probs}
p_{n+1}(v_L)  \propto \pi_{n+1}(v_L) \int \mathcal{K}(\bX_{n+1}; \btheta_{v_L}) p(\btheta_{v_L}\mid \mathcal{D}_n)\mathrm{d}\btheta_{v_L}, 
\end{equation}
where $\pi_{n+1}(v_L)$ is the prior probability defined in equation~\eqref{eq:priorProb}, while the integral is the posterior predictive distribution for DNA sequence $\bX_{n+1}$ under leaf $v_L$. Notice that if $v_L$ is a ``new'' taxon, equation~\eqref{eq:probs} becomes a prior predictive distribution. For the multinomial kernel with Dirichlet prior defined above, the integral is explicitly available. Specifically, it is straightforward to see that the marginal distribution for $X_{i,s}$ when the posterior follows a Dirichlet is 
\begin{equation}\label{eq:marginal_mult}
X_{n+1,s} \sim \textsc{mult}\Big(1; 
\frac{\xi_{v,s,\text{A}} + n_{v,s,\text{A}}}{M_{v,s}}, 
\frac{\xi_{v,s,\text{C}} + n_{v,s,\text{C}}}{M_{v,s}}, 
\frac{\xi_{v,s,\text{G}} + n_{v,s,\text{G}}}{M_{v,s}}, 
\frac{\xi_{v,s,\text{T}} + n_{v,s, \text{T}}}{M_{v,s}}\Big),
\end{equation}
where 
$M_{v,s} =\sum_{g \in \mathcal{N}_1} (\xi_{v,s,g} + n_{v,s,g})$ is a normalizing constant for the nucleotide probabilities. The corresponding prior predictive probability is obtained by setting $n_{v,s,g} = 0$ for every $g \in \mathcal{N}_1$ and normalizing by $\xi_{v,s,0} =\sum_{g \in \mathcal{N}_1} \xi_{v,s,g}$. Then, from equation~\eqref{eq:marginal_mult} and the location independence assumption, it can be easily shown that equation~\eqref{eq:probs} reduces to 
\begin{equation}\label{eq:final_probs}
    p_{n+1}(v_L) \propto 
    \begin{cases}
    \pi_{n+1}(v_L) \prod_{s=1}^p (\xi_{v_L,s, g_s} + n_{v_L,s,g_{s}})/M_{v_L,s}, & \textrm{if } v_L \textrm{ is an observed leaf}, \\
    \pi_{n+1}(v_L) \prod_{s=1}^p \xi_{v_L,s, g_s}/\xi_{v,s,0} , & \textrm{if } v_L \textrm{ is a novel leaf},
    \end{cases}
\end{equation}
where $g_s\in\mathcal{N}_1$ is the nucleotide of sequence $\mathbf{X}_{n+1}$ in locus $s = 1, \ldots, p$. Similar considerations hold for both the $\kappa$-product multinomial kernel and for the not aligned multinomial kernel. 

\subsection{Hyperparameter tuning}
From equation~\eqref{eq:final_probs}, it is straightforward to see that the hyperparameters $\bxi_{v,s}$ play an important role in defining the prediction probabilities. This is especially true for ``new'' leaves, since no nucleotides are observed. As already discussed in the main paper, uniform priors in this context may underestimate the predicted number of new taxa at the lowest level. Therefore, we need a method to tune $\bxi_{v,s}$ based on the information available in the taxonomic tree. To address this goal, we apply a method of moments estimator as detailed below.

For a node $v_\ell$ at level $\ell$, call $\mathcal{L}_n(v_\ell)$ the set of leaves linked to it. Under the assumption that
$$\btheta_{v,s} \stackrel{\text{iid}}{\sim} \textsc{dirichlet}(\xi_{v_\ell,s, \text{A}},\xi_{v_\ell,s, \text{C}}, \xi_{v_\ell,s, \text{G}}, \xi_{v_\ell,s, \text{T}}), \quad \text{for all } v \in \mathcal{L}_n(v_\ell),$$
each nucleotide probability is marginally distributed as $\theta_{v,s,g} \sim \textsc{beta}(\xi_{v_\ell,s,g}, \xi_{v_\ell,s,0}  - \xi_{v_\ell,s,g})$, with  $\xi_{v_\ell,s,0} = \sum_{g \in \mathcal{N}_1} \xi_{v_\ell,s,g}$ being the sum of the hyperparameters. From the moments of a beta distribution, it   holds that 
$$
E[\theta_{v,s,g}] = \frac{\xi_{v_\ell,s,g}}{\xi_{v_\ell,s,0}}, \qquad \text{and}  \qquad E[\theta_{v,s,g}^2] = \frac{\xi_{v_\ell,s,g}(\xi_{v_\ell,s,g}+1)}{\xi_{v_\ell,s,0}(\xi_{v_\ell,s,0}+1)},
$$
for $g\in \mathcal{N}_1$. Our goal is to estimate both $\xi_{v_\ell,s,0}$ and $\xi_{v_\ell,s,g}$. This can be done as follows. Recall that  $N_n(v)$ and $n_{v, s,g}$ are the number of sequences and the number of times nucleotide $g$ is recorded at locus $s$ for all sequences linked to leaf $v$, respectively. Our first method of the moments equation is
\begin{equation}\label{eq:mom_eq1}
\hat{\theta}_{v_\ell,s,g} = \frac{1}{N_n(v_\ell)}\sum_{v \in \mathcal{L}_n(v_\ell)}\frac{n_{v, s,g}}{N_n(v)} =  \frac{\xi_{v_\ell,s,g}}{\xi_{v_\ell,s,0}} = E[\theta_{v,s,g}].
\end{equation}
Here,  $\hat{\theta}_{v_\ell,s}^g$ is an average of the observed proportion of times nucleotide $g$ appears in the sequences linked to all leaves $v$ connected to $v_\ell$. For our second equation, we set

\begin{equation}\label{eq:mom_eq2}
\hat{S}_{v_\ell, s} = \frac{1}{N_n(v_\ell)}\sum_{v \in \mathcal{L}_n(v_\ell)}\sum_{g\in \mathcal{N}_1}\Big(\frac{n_{v, s,g}}{N_n(v)}\Big)^2 =   \sum_{g\in \mathcal{N}_1}\frac{\xi_{v_\ell,s,g}(\xi_{v_\ell,s,g}+1)}{\xi_{v_\ell,s,0}(\xi_{v_\ell,s,0}+1)} = \sum_{g\in \mathcal{N}_1}E[\theta_{v,s,g}^2],
\end{equation}
where $\hat{S}_{v_\ell, s}$ is the average sum of the squared nucleotide proportions for all $v$ linked to $v_\ell$. Notice that the third component in the equation can be further simplified as
$$
\sum_{g\in \mathcal{N}_1}E[\theta_{v,s,g}^2] = \sum_{g\in \mathcal{N}_1}\frac{\xi_{v_\ell,s,g}(\xi_{v_\ell,s,g}+1)}{\xi_{v_\ell,s,0}(\xi_{v_\ell,s,0}+1)} = \frac{1}{\xi_{v_\ell, s,0} + 1}\Big\{\xi_{v_\ell, s}^0\sum_{g\in \mathcal{N}_1}\Big(\frac{\xi_{v_\ell,s,g}}{\xi_{v_\ell,s,0}}\Big)^2 +1\Big\}.
$$
Then, plugging equation~\eqref{eq:mom_eq1} into \eqref{eq:mom_eq2} and letting $m_{v_{\ell}, s} = \sum_{g\in \mathcal{N}_1}\hat{\theta}_{v_\ell,s,g}^2$, one has that 
$$
\hat{S}_{v_\ell, s} = \frac{1}{\xi_{v_\ell, s, 0} + 1}(\xi_{v_\ell, s,0} m_{v_\ell, s} + 1),
$$
which, combined with equation~\eqref{eq:mom_eq1}, yields
\begin{equation}\label{eq:mom_estimators}
\xi_{v_\ell, s,0} = \frac{1-\hat{S}_{v_\ell, s}}{\hat{S}_{v_\ell, s} - m_{v_\ell, s}}, \qquad \text{and}  \qquad \xi_{v_\ell, s,g} = \xi_{v_\ell, s,0}\hat{\theta}_{v_\ell,s,g}, \quad g\in \mathcal{N}_1.
\end{equation}
The quantities in equation~\eqref{eq:mom_estimators} are the method of the moments estimators for our hyperparameters, and we can use them to borrow information across branches as discussed in the main paper. We detail the procedure in Algorithm~\ref{alg:mom} below.

\begin{algorithm}
\begin{algorithmic}[1]
\caption{Hyperparameter tuning via method of moments for the multinomial kernel}\label{alg:mom}
\For{leaf $v_L \in \mathcal{V}_L$}
\If{$v_L$ is a \textit{known} taxon}
    \State Estimate $\xi_{v_{L-1}, s, 0}$ and $\xi_{v_{L-1}, s,g}$, $g\in \mathcal{N}_1$, from equation~\eqref{eq:mom_estimators}, where $v_{L-1} = \text{pa}(v_L)$.
    \State Set prior $\btheta_{v_L, s} \sim \textsc{dirichlet}(\xi_{v_{L-1}, s, \text{A}},\xi_{v_{L-1}, s, \text{C}},\xi_{v_{L-1}, s, \text{G}},\xi_{v_{L-1}, s, \text{G}})$  
 \ElsIf{$v_L$ is a \textit{new} taxon}
    \State Estimate $\xi_{v_{\ell}, s,0}$ and $\xi_{v_{\ell}, s,g}$, $g\in \mathcal{N}_1$, from equation~\eqref{eq:mom_estimators}, where $v_{\ell}$ is the lowest possible \textit{known} taxon along the branch leading to $v_L$
    \State Set prior $\btheta_{v_L, s} \sim \textsc{dirichlet}(\xi_{v_{\ell}, s, \text{A}},\xi_{v_{\ell}, s, \text{C}},\xi_{v_{\ell}, s, \text{G}},\xi_{v_{\ell}, s, \text{G}})$  
\EndIf
\EndFor
\State Repeat procedure for all locations $s = 1, \ldots, p$.
\end{algorithmic}
\end{algorithm}

The idea behind Algorithm~\ref{alg:mom} is to incorporate taxonomic dependencies between the leaves, especially novel ones. The procedure works equally for the $\kappa$-product multinomial kernel.

\end{document}